\documentclass{article}

\usepackage{blindtext}
\usepackage{arxiv}

\usepackage[utf8]{inputenc} 
\usepackage[T1]{fontenc}    
\usepackage[hyphens]{url}
\usepackage{hyperref}       
\usepackage{booktabs}       
\usepackage{amsfonts}       
\usepackage{nicefrac}       
\usepackage{microtype}      
\usepackage{lipsum}
\usepackage{graphicx}
\usepackage{svg}
\usepackage[numbers]{natbib}
\usepackage{url}

\title{Medical Device Regulation Efforts for mHealth Apps - An Experience Report of Corona Check and Corona Health}

\author{
Marc Holfelder \\
LA2 GmbH\\
Erlangen, Germany\\
\texttt{marc.holfelder@la2.de}\\
\And
Lena Mulansky \\
Institute of Clinical Epidemiology and Biometry\\
University of Würzburg, Germany\\
\texttt{lena.mulansky@uni-wuerzburg.de} \\
\And
Winfried Schlee \\
Clinic and Policlinic for Psychiatry and Psychotherapy\\
University of Regensburg, Germany\\
\texttt{winfried.schlee@gmail.com} \\
\And
Harald Baumeister\\
Department of Clinical Psychology and Psychotherapy\\
Ulm University, Germany\\
\texttt{harald.baumeister@uni-ulm.de} \\  
\And
Johannes Schobel\\
Institute DigiHealth\\
Neu-Ulm University of Applied Sciences, Germany\\
\texttt{johannes.schobel@hnu.de} \\  
\And
Helmut Greger \\
Service Center Medical Informatics\\
Würzburg University Hospital, Germany\\
\texttt{Greger\_H@ukw.de} \\
\And
Andreas Hoff \\
LA2 GmbH\\
Erlangen, Germany\\
\texttt{andreas.hoff@la2.de} \\
\And
Rüdiger Pryss \\
Institute of Clinical Epidemiology and Biometry\\
University of Würzburg, Germany\\
\texttt{ruediger.pryss@uni-wuerzburg.de} \\
}

\begin{document}
\maketitle

\begin{abstract}
Within the healthcare environment, mobile health (mHealth) applications (apps) are more and more important. The number of new mHealth apps has risen steadily in the last years. Especially the Covid-19 pandemic has led to an enormous amount of app releases. Notably, in most countries, mHealth applications have to be already compliant with several regulatory aspects in order to be declared to be a 'medical app'. However, the latest applicable medical device regulation (MDR) does not comment in more detail on the topic of the requirements for mHealth applications. When developing a medical app, it is essential that all contributors in an interdisciplinary team - especially the software engineers - are aware of the specific regulatory requirements beforehand. The development process, however, should not be stalled too long due to the integration of the MDR. Therefore, a developing framework, which includes these aspects, is required, to enable a smooth development process. The paper at hand introduces the creation of such a framework on the basis of the Corona Health and Corona Check apps. The relevant regulatory guidelines are listed and summarized to a guidance for medical app developments. In particular, the important stages and faced challenges emerged during the entire development process are highlighted.
\end{abstract}

\keywords{mHealth \and mobile application \and MDR \and medical device regulation \and medical device software}

\section{Introduction}\label{sec:introduction}
Regardless of pandemic mobile apps, regulatory requirements in the area of medical mobile apps have been neglected for a very long time. In 2016, 2.5 million medical apps are listed by Statista in the two prominent app stores of Apple and Google \cite{Statista2016}. The number is increasing almost daily by orders of magnitude that would have been unthinkable 10 years ago \cite{StatistaGoogle}\cite{StatistaApple}. Since regulatory frameworks for medical apps have not yet been the focus of governmental audits, anyone can deploy a medical app to the mentioned stores without much responsibility. This causes two main problems:

\begin{itemize}
    \item Initially, little can be said about whether a selected app actually helps or is effective with limited overall quality in most health promotion and health care domains \cite{portenhauser2021,sander2020,schultchen2020, stach2020,terhorst2018, terhorst2021}. Regulatory requirements cannot provide conclusive support in this regard, but they would nevertheless give an indication of whether an app in question adheres to standards regarding validation, verification, and the fundamentals of data integrity.
    \item The second problem is that, given the vast number of medical apps on offer, patients cannot know which app is actually the right one for their need at hand. Regulatory frameworks, including procedural quality assurance, can be a quality feature here, and at the same time, also a clue to enable better navigation for those affected and offer a more reliable assessment of an app.
\end{itemize}

Two medical apps that adhere to the same specifications are therefore easier to compare. Hence, their mode of action can also be better compared. Fortunately, both legislators and many initiatives (VDI, DIN, DKE, etc.) are increasingly taking care by creating regulatory frameworks for medical apps. There are also an increasing number of initiatives working to provide navigation assistance for medical apps, such as MHAD \cite{stach2020} or AppRadar \cite{AppRadar}. However, because such initiatives and guidelines have only recently begun to exist, the app market (both in research and academia, as well as in the medical usage space) has only rarely implemented these guidelines yet. This could also be due to the fact that too few of these guidelines are yet actually used and enforced by the relevant authority representatives as a basis for their audits.

Regulatory requirements must not be a stumbling block, because speed is another very important factor in the development of future apps, together with reliability, costs, and acceptance. It is not worthwhile if an app works well and is accepted, but any regulations and their application is accompanied by the fact that the app is only ready for use when the actual pandemic has already subsided.
The regulatory requirements must be formulated in such a way that they are understandable and do not interfere with the upcoming development work, but rather support it. In the following, a field report 
of developing the two mobile apps Corona Check \cite{CoronaC} and Corona Health \cite{CoronaH} is presented, including a derived realization concept on how these apps could be developed quickly, cost-effectively, and tailored to pandemic needs. At the same time, a high level of quality had to be achieved in order to generate the greatest possible acceptance among the population.

Corona Check was developed within the framework of a scientific collaboration between German university partners, the Bavarian Health Authority, and software companies. It mainly provides 7 questions to help with the early detection of COVID-19. Corona Check thereby incorporates the ideas of patient reported outcome and mobile sensing, combined with a direct feedback to users by the app based on the answers of the 7 mentioned questions. The overall goal was to provide a quick test that can be easily performed at any time and at any change of symptoms.

Corona Health, in turn, was also developed within a broader framework of participants and provides studies that are used to evaluate the impact of the Covid-19 pandemic on mental and physical health. The gathered results should show the need for improvements in the health systems as well as pertaining to the individual coping skills. The outcome should also be used for recommendations to decrease negative impacts, which arise through the pandemic itself as well as through the steps to curb it. Corona Health is based on the following principles: patient reported outcomes, ecological momentary assessments, mobile crowdsensing as well as sensor measurements. The latter encompass GPS measurements as well as data on the usage of apps like Facebook, only if users allow these measurements. For a deeper understanding of these concepts, see \cite{kraft2020combining,pryss2019mobile}.

When developing these two apps, the requirement arose to adhere to the Medical Device Regulation (MDR). Both apps are released into the official app stores from Google and Apple and are compliant with the MDR. To achieve the objective of compliance, the following aims were created by the teams that developed these apps:

\begin{itemize}
    \item Establish a harmonized regulatory approach that draws the best from all rules and standards, thus avoiding "over-regulation", but still showing conformity with the Medical Devices Regulation.
    \item Possibility of an iterative development approach while maintaining the necessary regulations.
    \item Creation of a simple and feasible procedure, so that the high development speed is not slowed down, but at the same time all necessary requirements are implemented in an app solution.
\end{itemize}

And as a kind of side effect - the creation of an approach through which efficient error prevention becomes possible due to a more detailed requirements phase:
\begin{itemize}
    \item Reliable project schedules can be created with reliable cost estimates.
    \item Reduction of error costs (i.e., a detailed requirements phase prevents errors in testing or after going live).
\end{itemize}

The important aspects to achieve the aforementioned objectives are discussed in this paper. We hope to help other mobile app developers in creating mHealth apps in the context of regulatory needs. The reminder of this paper is organized as follows. In Section \ref{sec:relatedwork}, relevant work and existing standards, which are useful and suitable as a regulatory basis for the development of medical apps - in general and for a pandemic app in particular - are presented. The creation/implementation of a pragmatic procedural model is the basis for the provision of correctly created regulatory apps and thus gaining the trust of citizens. Section \ref{sec:framework} includes the detailed concept of a regulatory compliant app creation process, whereas Section \ref{sec:discussion} exemplify this process on the basis of two realized application development projects. The conclusion of the paper, consisting of a summary and an outlook, is covered in Section \ref{sec:sando}.

\section{Background and Related Work}\label{sec:relatedwork}
The Medical Device Directive (MDD) 93/42/EEC \cite{CouncilDirective} has been revised and substituted by the Medical Device Regulation (MDR[EU)] 2017/745 in 2017 \cite{Regulation}. In \cite{MDRKeutzer} and \cite{MDD}, the authors present the new Medical Device Regulation and its key elements. The author of \cite{MDD} points out the differences between the new and the foregone regulation, whereas in \cite{MDRKeutzer}, the elements of the new MDR are revealed. Particularly, the workflow and important steps of the technical documentation during the app development process are examined. Moreover, approaches to classify software and applications in respect of risk classes are shown \cite{MDRKeutzer}. Trektere et al. \cite{MDevSpiceTrektere},  \cite{TracebilityTrektere} point out the necessity for a mHealth application development framework and its key criteria. They present an existing framework for medical device software development, and an approach, how it can be tailored to be used in the context of mHealth application development. In \cite{TracebilityTrektere}, the author put particular attention on the key criteria "Traceability", which is essential in the development process to ensure a reliable software \cite{TracebilityTrektere}.

Since neither the Medical Devices Act nor the Medical Device Regulation comment in more detail on the topic of requirements for a medical app, at first, common standards and regulations had to be identified \cite{CouncilDirective}\cite{Regulation}. Further, they had to be analyzed and compared with regards to relevant topics. From this, a suitable and harmonized regulatory basis was created, on the basis of which project managers, developers, and quality managers could jointly develop an app that was complete, (almost) error-free and of high quality. The following regulations from the healthcare sector were analyzed:

\underline{Collection of common standards in the healthcare sector}
\begin{itemize}
    \item IEC 62304 - Medical devices-Software - Software life cycle processes\\
    $\rightarrow$ The IEC 62304 standard specifies the requirements for software life cycle processes for the development of medical software and software in medical devices. The processes, activities and tasks prescribed in this standard form a common framework for action for all life cycle processes for software in the field of medical devices.\cite{IEC62304}
    \item GAMP5 - "Good Automated Manufacturing Practice Supplier Guide for Validation of Automated Systems in Pharmaceutical Manufacture"\\
    $\rightarrow$ This guide has become the standard set of rules for the validation of computerized systems in the pharmaceutical industry (manufacturers and suppliers). However, the GAMP5 set of rules is not legally binding. Therefore, different forms of validation of computerized systems are possible, which is useful for many systems.\cite{GAMP5}
    \item General Principles of Software Validation - Regulations of the U.S. Food and Drug Administration (FDA) - U.S. agency for food and drugs.\\
    $\rightarrow$ This guidance outlines general validation principles that the FDA considers applicable to the validation of medical device software or the validation of software used in the design, development, or manufacture of medical devices.\cite{FDA}
    \item PICS 11-3 PIC/S Guide to Good Manufacturing Practices (GMP)\\
    $\rightarrow$ The PIC/S Guide to GMP is the basis for GMP inspections. In particular, its Annex 11, "Computerized Systems," is used in the inspection of such systems.The purpose of this document is to provide recommendations and background information on computerized systems to assist inspectors in training and during the inspection of computerized systems. The document is intended to serve as "good practice" for inspectors that are responsible for inspecting applications in the regulated pharmaceutical sector.\cite{PIC}
    \item IEC 82304 Health software - Part 1: General requirements for product safety.\\
    $\rightarrow$ This standard covers "software as a medical device", including Mobile Medical Apps as well as other Health Software (outside the application of Regulation (EU) 2017/745 on medical devices). IEC 82304-1 has closed a gap in the normative coverage of validation and documentation of software placed on the market without specific hardware. The key points of this standard are requirements for product safety (SAFETY) as well as for information security (SECURITY) of health software products. This also includes requirements for usability (USABILITY) and instructions for use (INSTRUCTION FOR USE). The standard requires compliance with a software development process and refers to IEC 62304/A1:2015.\cite{IEC82304}
\end{itemize}

\underline{Comparison of Standards}\\
\\
After the successful collection of the required regulations, a topic catalog was created by the team of Corona Check and Corona Health, in which all important topics and regulatory requirements on the topic of app development were collected. The following points were agreed upon: Software life cycle with the topics validation plan/documentation (incl. verification), risk management as well as the involvement of service providers and the training of employees. Subsequently, it was examined whether the above-mentioned main topics were addressed by the various sets of rules. If this was the case for a set of rules, then the requirements of the respective key topic in the relevant set of rules were checked for the scope and proportionality of the activities required there.
In summary, all of the aforementioned sets of rules addressed the specified topics and were each very similar in the scope and proportionality of the activities required therein.
Table \ref{table:standards} is intended to provide a general overview.

\begin{table}[ht]
  \centering
  \includegraphics[width=1.0\linewidth,keepaspectratio]{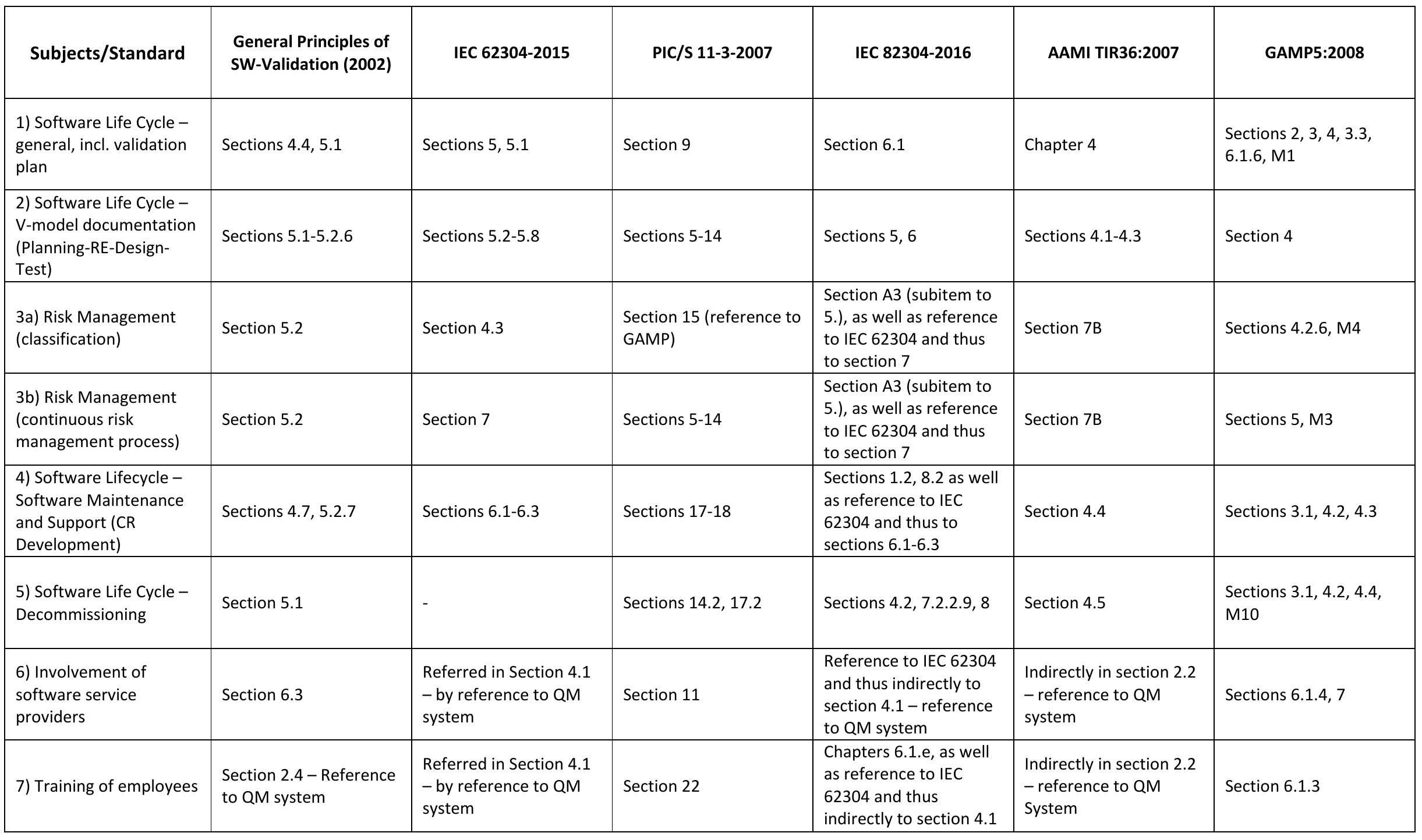}
  \caption{Comparison of the relevant regulatory standards}
  \label{table:standards}
\end{table}

The previously defined main topics were also adhered to after completion of the analysis. After further investigation, it was decided that with IEC 82304, a set of rules could be used that was designed precisely for such a case (standalone software systems and health apps). However, a major problem was that the requirements of this standard still had quite extensive consequences on the scope of possible documentation and the development speed of an app project.\\
Essential points from the other standards were therefore updated in the planned development process, which was then additionally "tailored" in a risk-based approach, i.e., finalized (one could also say "streamlined"). In this way, the other standards were taken into account and, despite this, the regulatory requirements for an app development project could be reduced to a sensible minimum level of effort. In other words, the initial development process included the "good stuff" from all of the previously outlined standards and still contains all of the relevant specifications of IEC 82304-1. This concrete conceptual procedure is described in detail in Section 3 "Concept of the regulatory compliant app creation process". It can be used as a template for upcoming projects. It should also be noted that compliance with this regulatory-compliant development process was ultimately essential for the medical apps developed to be included in the Google (Google Play) and Apple (App Store) stores. However, this did not take the form of sending a release certificate directly to Google or Apple, but this release certificate in turn formed the basis for
\begin{itemize}
    \item on the one hand, a positive vote by an ethics committee (which ultimately examines the content according to ethical, medical, and scientific aspects, among others),
    \item and on the other hand, the support of a public body (Robert Koch - Institute (RKI), Bavarian State Office for Health and Food Safety, ...)
\end{itemize}
to be maintained. Both were ultimately key conditions for store operators to include a Covid-19 medical app as such in the stores. However, in order to obtain the final approval of the ethics committee and the public body, in addition to a release certificate, further suitable proof had to be provided that the aforementioned standard(s) - including the harmonized cut - were actually complied within a software project. This was done in the form of so-called validation documents, training certificates, and Standard Operating Procedures (SOPs). They had to be submitted to these bodies as "proof" of compliance with Medical Device Regulation/Ordinance on Medical Device specifications or their (above-mentioned) underlying standards. The following chapter describes how this process was structured, which documents were required, and what their basic contents are.

\section{Concept of the regulatory compliant app creation process}\label{sec:framework}
Proof that regulatory requirements have been met in an app development project is provided by supplying the documents mentioned in the previous chapter. In order for these to be provided properly, it is necessary for the development project to be planned in a regulatory manner. Since all the regulations mentioned here recommend a risk-based approach, this also forms the basis for the further procedure in planning. More details can be found in the section "Risk management process/analysis". 

\subsection{Software life cycle requirements} 
During planning, in addition to the development phases and the corresponding documents (e.g., requirements phase = specifications, acceptance test = system test, etc.), service providers, training courses, test concept, and the procedure for requirement tracing (ensuring that all requirements are actually contained in the software at the end) must be defined. Before this, however, the roles, their responsibilities in the project, and their assignments to the current project members must be defined at first.

\subsubsection{Roles and assignment to project team members} 
At least, the following roles should exist in a software development project:
\begin{itemize}
    \item Project Management:\\
    Responsible for the implementation and management of the entire product development.
    \item Quality Management:\\
    Responsible for ensuring software quality, regulatory requirements as well as the archiving of documents.
    \item Development:\\
    Responsible for development planning as well as related testing (except system testing).
    \item Test:\\
    Responsible for overall system testing (planning and execution) and risk management. \\ Note: Risk management may be a separate role.
    \item Main responsible person:\\
    Main responsible for the whole project. Most importantly, issuance of the final release
\end{itemize}

Current team members must be assigned to their respective roles in the project. Necessary trainings still have to be defined (see also the point "Involved service providers, trainings and necessary SOPs" below). Once all this has been defined, the next step is to define the development phases (including the corresponding documents) and the role responsibilities for documents as well as the review planning.

\subsubsection{Development phases/documents, responsibilities and reviews}
Since app development processes usually have a manageable complexity, development steps required by regulations can usually be combined during planning. 
A harmonized approach from IEC 62304 Class B and GAMP5 Cat. 4 (configurable software) was used as the basis for further planning of the development process.
Accordingly, the development phases of the project with the associated documents are structured as follows:
\begin{itemize}
    \item Planning phase (document title: Software Quality Management Plan or also Master Validation Plan)
    \item Requirements elicitation (document title: requirements specification or requirements specification)
    \item Summarized functional description + design (document title: combined functional specification/design specification)
    \item Programming (document title: no separate document is created here)
    \item Combined module/integration test (document title: Combined module/integration test with specification/protocol/evaluation)
    \item System test (document title: System test with specification/protocol/evaluation)
    \item Release (document title: release protocol, e.g., for ethics committee and store operator)
\end{itemize}

The commonly known waterfall process (here, the so-called procedure model of software development or also called V-model), with the corresponding documents, would usually be like it can be seen in Figure \ref{fig:v-model}:

\begin{figure}[htpb]
  \centering
  \includegraphics[width=0.6\linewidth,keepaspectratio]{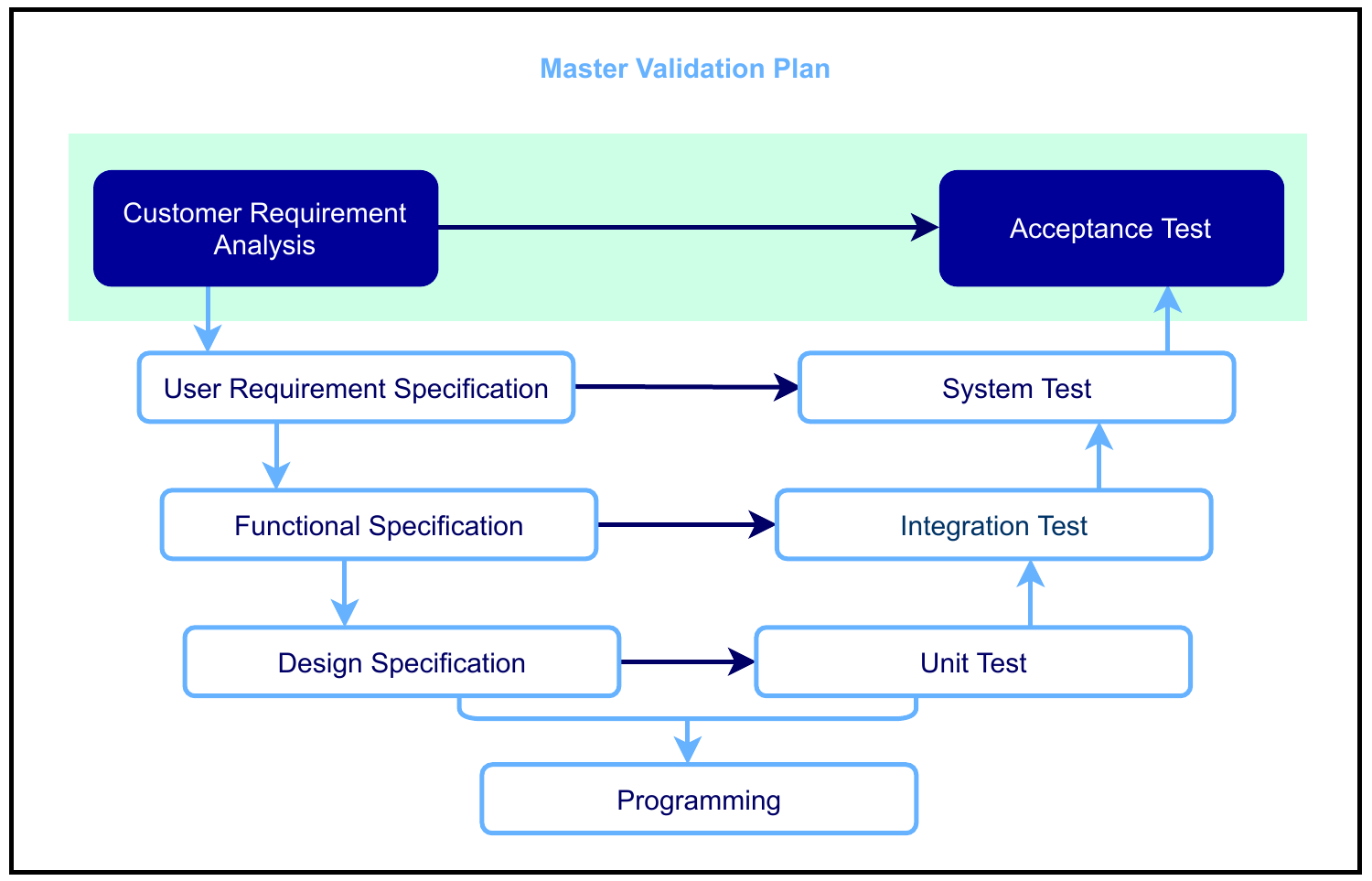}
  \caption{Waterfall (V-model) software development process model}
  \label{fig:v-model}
\end{figure}

\begin{figure}[htpb]
  \centering
  \includegraphics[width=0.6\linewidth,keepaspectratio]{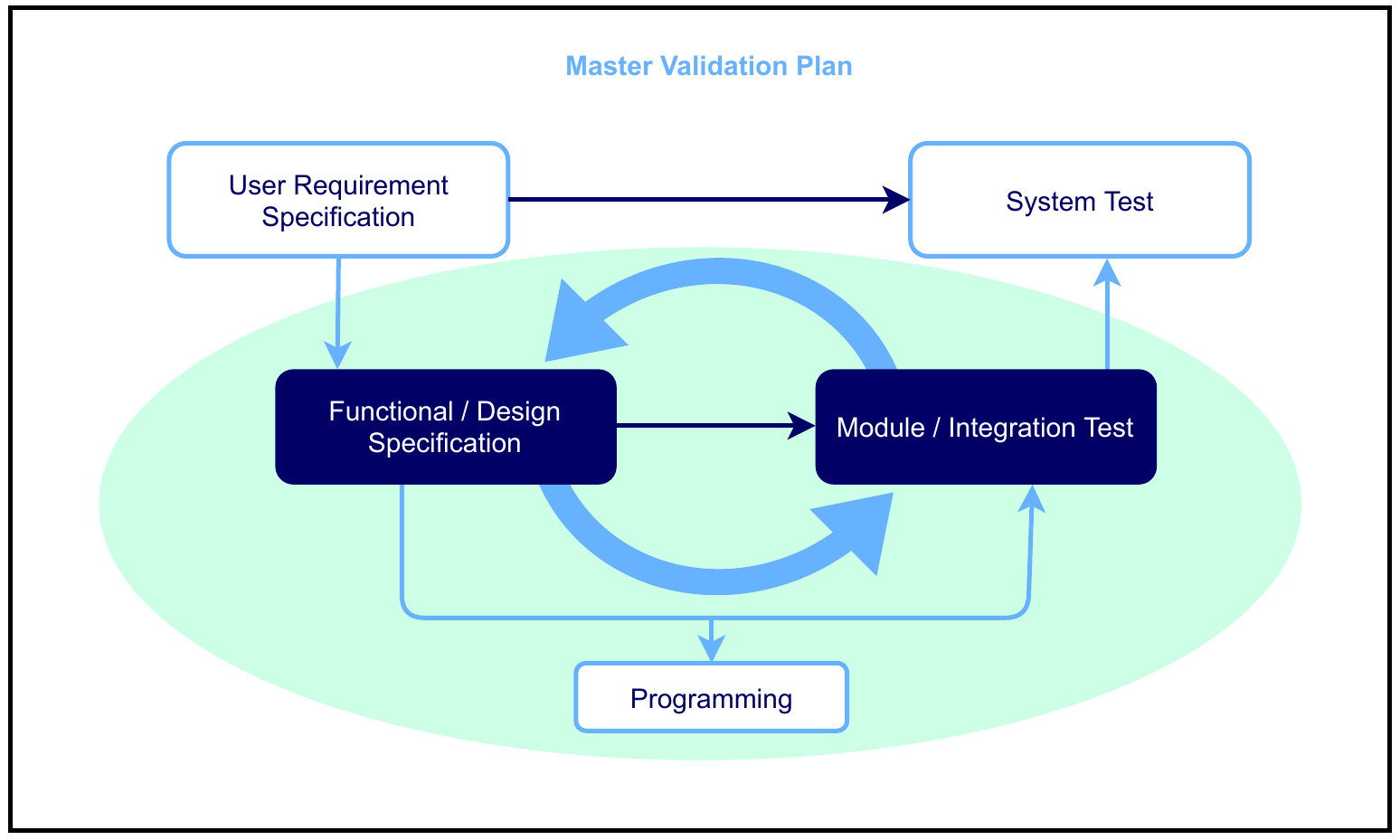}
  \caption{Process model of agile software development}
  \label{fig:Agil-model}
\end{figure}

However, since we recommend an agile software development process for (development) speed reasons, the actual development process is planned as shown in Figure \ref{fig:Agil-model}.

In the planning, it is defined that the requirements must be firstly collected in detail in order to have the basis (the so-called backlog) to implement these requirements in an agile process "design - programming - test" (green-shaded area). Once all requirements have been implemented and tested in the various sprints (short development cycles), they must be finally tested in the system test. All validation and verification documents can be left in the design status until the final release is due. Interim test results after each sprint are documented. Towards the end of the project, these documents can be released step by step. Now that roles, development process, and responsibilities have been defined, the requirements for service providers, training and necessary SOPs must be formulated.

\subsubsection{Involved service providers, training and necessary SOPs}
The service providers involved are listed, their qualification and availability requirements formulated, and any organizational framework conditions defined. Before the start of the project, the project manager must ensure that the employees involved (both his own and those of the service provider) have sufficient experience and are trained in the regulatory principles relating to app development/validation. Likewise, all training required of the respective project members (according to their roles) should be defined. Among other things, the team members should have read and understood the procedure described here in the Master Validation Plan (or also called Software Quality Management Plan). Results must be documented in a log file or in training cards/passes. In addition, all SOPs necessary for the use and operation of an app are presented. Once all this has been done, it must now be ensured that the app actually meets all the required specifications. The verification is done formally in a so-called requirement tracing process.

\subsubsection{Tracing processes}
It must be planned that the requirement tracing (tracing of requirements across the development process) must be carried out in three passes. This ensures the presence of all loads in the combined requirements specification/design specification and at least one test in each of the respective test specifications (module/integration test specification - system test specification). The following documents are therefore created:
\begin{itemize}
    \item Trace matrix Requirements specification - Requirements specification/design specification
    \item Trace Matrix Specifications - System Test Specification
    \item Trace matrix Requirements/design specification - Module/integration test specification
\end{itemize}

After the requirements for the tracing process have been completed, last but not least, a lean test concept must be formulated. 

\subsubsection{Test concept}
The tests to be performed have already been planned in the development process and the underlying documents named. The module/integration tests can take place in a simulated environment (it is recommended here to test in the productive environment as soon as possible). The final system tests must take place in a productive environment. The tests are usually carried out in 2 phases. In phase 1, the so-called module/integration tests have to be passed, and, in phase 2, the system tests (or also called acceptance tests). In this process, the module/integration tests are always completely defined (and released) per iteration, executed, and the test results are finally logged and evaluated. If complete tests cannot be performed in an iteration, the regression tests to be performed must be defined in advance. The printed test specification represents the test protocol, which must be completed by the testers. After completion of the tests, each tester checks the completeness and validity of the entries and confirms this by date, abbreviation, name, and his signature on the printed test specification for each of the requirements tested by him/her. Alternatively, the assignment of signature and abbreviation can also be recorded once in tabulator form, in which case it would be sufficient for the tester to confirm the completeness and validity of the entries by date and abbreviation per test case. Note: Where appropriate, the test result should be supported by one or more screenshots. The test results are then to be evaluated by the respective responsible persons. This can happen in an additional document. Once the successful system tests have been completed, the software can be released. (see Section \ref{subsec:release}).

\subsubsection{Interfaces}
Not every app has interfaces to other systems. However, if this is the case, then these interfaces must also be "validated and verified". As an exception, this is now done on the basis of the GAMP5 set of rules. The procedure described there for "Category 3: Non-configured product" is used here. This means that requirements must be formulated for the relevant interface and the corresponding system test cases. The documentation can be done in a single document, a so-called Project Summary (also called "Fact File" or "Project Profile"). This document contains a short organizational section, requirements specification, acceptance test specification including protocol, traceability matrix, and risk management elements. In terms of content, the data to be imported (data structure) must be defined, the transfer type/route (including secure/near-time data transfer, interpretation logic, and configuration) must be specified, and the data storage locations (data transfer or storage) must be formulated.

\subsection{Release}\label{subsec:release}
After completion of the training courses and creation of the aforementioned documents, including successful system tests and the formulation of any necessary SOPs and project summaries for interfaces, the app can be released: For this purpose, an official release document must be created as proof for authorities. It must contain a list of all created and released (verification) documents and be signed by all project participants with primary responsibility.

\subsection{Risk management process/analysis}
The risk management process taking place for a development project in the regulated environment should always happen at least on the basis of ICH Q9 Quality Risk Management (ICH = International Council for Harmonization of Technical Requirements for Pharmaceuticals for Human Use) \cite{ICH}. In this process, at the beginning of the project and during the development process, risks are constantly
\begin{itemize}
    \item collected
    \item rated
    \item mitigated and
    \item monitored.
\end{itemize}

Note: The process shown here is simplified for explanatory purposes. Risks have to be divided into at least into one of the following groups:

\begin{itemize}
    \item Product risks
    \item Patient risks
    \item Risks related to data integrity
\end{itemize}

All stakeholders should be represented in the risk management process. These can be, for example, project managers, developers, quality manager, but also users, representatives of the authorities, or people with IT development experience. If these risks cannot be mitigated by the IT, efforts can also be made to reduce the risks to a necessary minimum through training and detailed SOPs (Standard Operational Procedures). Since the actual risks can vary from app to app and situation to situation, no presentation or summary of possible risks was provided. However, to give an impression of the effectiveness and efficiency of the risk management process, its results in the Corona Health project have been summarized in Figure \ref{fig:risk-matrix} as an example:

\begin{figure}[htpb]
  \centering
  \includegraphics[width=1.0\linewidth,keepaspectratio]{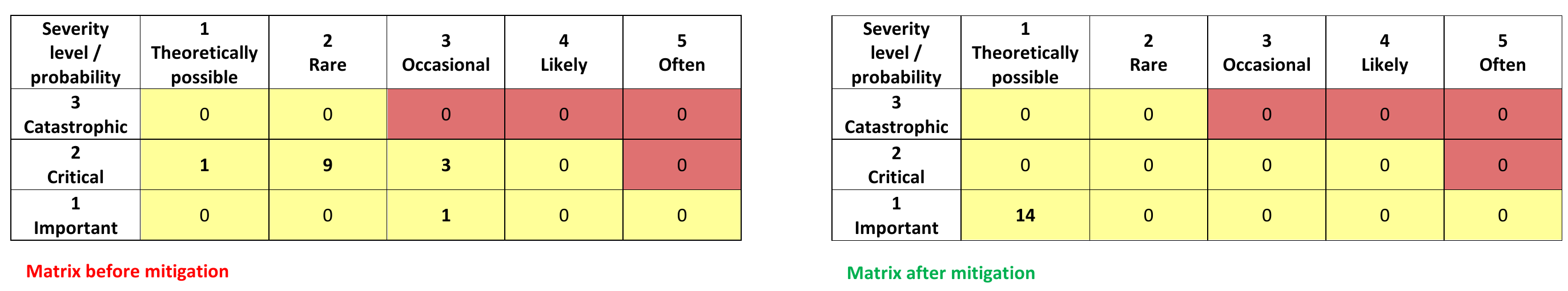}
  \caption{Results of the risk management process in the Corona Health project}
  \label{fig:risk-matrix}
\end{figure}

The left table shows the classification of the risks before mitigation, the right table after risk mitigation. Care must be taken to ensure that none of the risks fall into the red defined range (intolerable range) and that the risks are reduced as far as possible to a tolerable minimum. Risk management should be an essential part of future IT development projects for medical apps. Having so far described the planning of the regulatory-compliant software development project and the risk management process, the following chapters will show how thorough risk-based planning had an impact on the project progress and success of the apps Corona Check and Corona Health.

\section{Discussion}\label{sec:discussion}
In 2020, the mobile apps Corona Check and Corona Health were consecutively created. Both projects went through the planning and software development process described above. Sequences and internal durations were very similar in both projects (depending on the project scope). External factors were responsible for the different overall project durations. The first thing to mention here are the releases in the Apple and Google stores, which took about 11-50 days. In general, how can the differences be explained? Between 03/2020 - 08/2020, Google and Apple changed their store approval conditions. Initially, the two companies did not set any requirements for so-called Corona apps, but this changed after a few weeks, and the vote of an ethics committee became mandatory. After another 3-4 months, the recommendation of an official body (for example, for Corona Health: the RKI) was required as well. Further, the additional time through multiple internal release rounds of the store operators (duration about 5-6 days each) subsequent to the fulfillment of the requirements were among the external factors. Moreover, the subsequent approval of central search terms such as "Corona" and "Covid" took at least 45 days for the Google Play store. More specifically, the latter period of time had to pass until Google found the apps when using these search terms. For example, from the approval by the ethics committee to the moment Corona Health was found in the Google Play Store, under the search terms "Corona" or "Covid", 82 days passed.  Why is the release for central search terms so important? If an app does \underline{not} appear in the hit list for a general search term such as "Corona", it is practically not found at all in the stores. In our case, the search term "Corona Health App" had to be entered into the Google Play Store for the Corona Health app to be found and displayed at all. When the app was finally unlocked for the central search terms, the download numbers multiplied by orders of magnitude. However, it must be also clearly articulated that in the case of the Google Play store, it was and still is a courtesy of Google to maintain a special Covid-19 search realm. Apps that are listed in this realm are particularly displayed. The mentioned waiting time for being listed in the Google Play Store is therefore related to this particular search realm. It is reasonable that Google evaluates mobile apps in a time like a pandemic in an in-depth manner. In the case of the Apple app store, no particular realms were offered. 

\begin{figure}[htpb]
  \centering
  \includegraphics[width=0.8\linewidth,keepaspectratio]{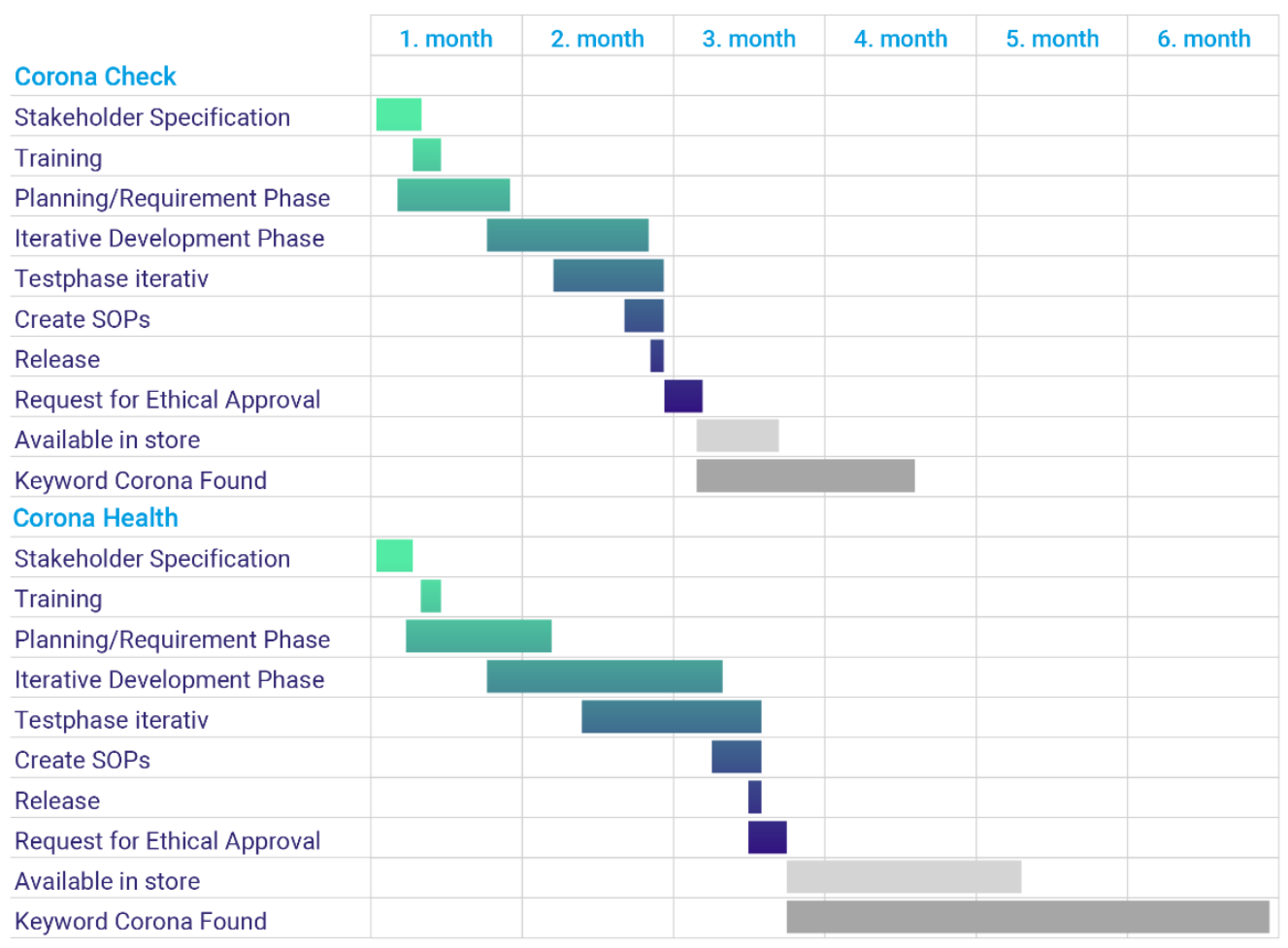}
  \caption{Development times for Corona Check and Corona Health}
  \label{fig:ganttchart}
\end{figure}

Figure \ref{fig:ganttchart} summarizes relevant time aspects. It is easy to see that the difference in the development times of the two app projects is only about two weeks. If we solely rely on the development period, the following schedule results on the basis of these two projects for an app with medium scope and complexity (the company selection must have already taken place here):

\begin{itemize}
    \item Determination of the project participants (Day 1 - Day 8)
    \item Training Phase (Day 9 - Day 11)
    \item Planning Phase and Requirement Phase (Day 8 - Day 35) 
    \item Development phase (Day 30 - Day 72) 
    \item Test phase (Day 44 - Day 79)
    \item Create and release SOPs (Day 70 - Day 79)
    \item Release (Day 79)
    \item Request Ethics Committees (Day 79 - Day 86)
    \item Release by the store operators (Day 86 - Day?)
    \item Found on Google under the search term Corona (Day 86 - Day?)
\end{itemize}

In the project sequence shown here, it is noteworthy that the relatively long requirements phase and regulatory framework hardly slow down the actual development speed. On the contrary, the relatively long requirements phase and regulatory specifications mean that ambiguities and problems can be identified and eliminated in a reasonable amount of time, including that potential errors (sources) are identified and eliminated in a rather short period of time. Thus, a solid, consistent, testable, and complete requirement documentation is created on the basis of which the following can be achieved: Short development cycles, combined with extensive testing and fast feedback loops, should be aimed at, so that the app development can be completed in 3 iterations, in the best case. The validation documentation then helps with further developments of the apps, additionally shortening the times to onboard and train project members in question. Existing tests can be planned in detail and, if necessary, expanded through existing extensive test specifications, thus forming the basis for finally performing the tests successfully \cite{Vogel2021MobileSoft}.

\section{Summary and Outlook}\label{sec:sando}
Based on years of experience in the field of app development as well as regulatory software development in general, those which were responsible for the Corona Health / Corona Check projects were able to reduce the actual development times to a minimum (see above), while still achieving proper development costs. The Corona -Warn app, which is more complex in terms of its efforts, had a realization time of roughly 50 days, but without regulatory compliance. We estimate future app development projects as follows: With proper planning and taking the regulatory compliance as well as a feasible requirement engineering process into account, it can be reliably estimated that the development time will be extended by a maximum of approximately 10\% due to compliance with regulatory requirements. Experiences show that the collaboration of experienced app developers, good RE experts, and regulatory experienced and pragmatic quality managers can even accelerate development times, while preserving product quality on the other hand. To conclude, the following objectives, set by the project management at the beginning of both projects, could be achieved by the overall app team:

\begin{itemize}
    \item Conformity with the Medical Devices Regulation 
    \item Consensus on iterative development approach, including ...
    \item ... high development speed,
    \item ... feasible project schedules, and
    \item ... low overall project costs.
\end{itemize}

In total, the two Corona apps were created each in less than 4 months at proper costs, including approvals from German ethic committees. An early contact with Google and Apple in order to be quickly included in the stores, on the one hand, and to be found well by potential users on the other hand is the only point that could not be planned sufficiently at the beginning of both projects. If official bodies (such as the RKI or the Federal Ministry of Health) can even be involved in the development process, then, novel medical apps can be quickly made available for a pandemic. This includes the provision of a high quality, the compliance with regulatory aspects as well as achieve acceptable costs.

\section*{Acknowledgments}\label{sec:acks}
We thank Dr. Caroline Cohrdes (Robert Koch-Institute) and Johanna-Sophie Edler (Robert Koch-Institute) for their support; we further thank Carsten Vogel (University of Würzburg) as well as Fabian and Julian Haug for their development of the mobile applications. We also thank all those, who tested the applications. 
The paper at hand was created within the COMPASS project, which is part of the German COVID-19 Research Network of University Medicine (“Netzwerk Universitätsmedizin”), funded by the German Federal Ministry of Education and Research (funding reference 01KX2021).
Although this work was supported by the Robert Koch-Institute (RKI), it solely reflects the author's view and does not necessarily reflect the views of the Robert Koch-Institute or any other organization.
\bibliographystyle{unsrtnat}

\begin{thebibliography}{27}
\providecommand{\natexlab}[1]{#1}
\providecommand{\url}[1]{\texttt{#1}}
\expandafter\ifx\csname urlstyle\endcsname\relax
  \providecommand{\doi}[1]{doi: #1}\else
  \providecommand{\doi}{doi: \begingroup \urlstyle{rm}\Url}\fi

\bibitem[Statista(2016)]{Statista2016}
Statista.
\newblock {Anzahl der verfügbaren Gesundheits-Apps nach App Store im Jahr
  2016}.
\newblock
  \url{https://de.statista.com/statistik/daten/studie/688545/umfrage/anzahl-der-verfuegbaren-gesundheits-apps-nach-app-store/},
  2016.
\newblock Accessed: 2021-03-30.

\bibitem[Statista(2021{\natexlab{a}})]{StatistaGoogle}
Statista.
\newblock {Number of mHealth apps available at Google Play from 1st quarter
  2015 to 4th quarter 2020}.
\newblock
  \url{https://www.statista.com/statistics/779919/health-apps-available-google-play-worldwide//},
  2021{\natexlab{a}}.
\newblock Accessed: 2021-04-01.

\bibitem[Statista(2021{\natexlab{b}})]{StatistaApple}
Statista.
\newblock {Number of mHealth apps available in the Apple App Store from 1st
  quarter 2015 to 4th quarter 2020}.
\newblock
  \url{https://www.statista.com/statistics/779910/health-apps-available-ios-worldwide/},
  2021{\natexlab{b}}.
\newblock Accessed: 2021-04-01.

\bibitem[Portenhauser et~al.(2021)Portenhauser, Terhorst, Schultchen, Sander,
  Denkinger, Stach, Waldherr, Dallmeier, Baumeister, and
  Messner]{portenhauser2021}
Alexandra~A Portenhauser, Yannik Terhorst, Dana Schultchen, Lasse~B Sander,
  Michael~D Denkinger, Michael Stach, Natalie Waldherr, Dhayana Dallmeier,
  Harald Baumeister, and Eva-Maria Messner.
\newblock Mobile apps for older adults: Systematic search and evaluation within
  online stores.
\newblock \emph{JMIR Aging}, 4:\penalty0 e23313, 2021.

\bibitem[Sander et~al.(2020)Sander, Schorndanner, Terhorst, Spanhel, Pryss,
  Baumeister, and Messner]{sander2020}
Lasse~Bosse Sander, Johanna Schorndanner, Yannik Terhorst, Kerstin Spanhel,
  R{\"u}diger Pryss, Harald Baumeister, and Eva-Maria Messner.
\newblock Help for trauma from the app stores?’ a systematic review and
  standardised rating of apps for post-traumatic stress disorder (ptsd).
\newblock \emph{European Journal of Psychotraumatology}, 11:\penalty0 1701788,
  2020.

\bibitem[Schultchen et~al.(2020)Schultchen, Terhorst, Holderied, Stach,
  Messner, Baumeister, and Sander]{schultchen2020}
Dana Schultchen, Yannik Terhorst, Tanja Holderied, Michael Stach, Eva-Maria
  Messner, Harald Baumeister, and Lasse~Bosse Sander.
\newblock Stay present with your phone: A systematic review and standardized
  rating of mindfulness apps in european app stores.
\newblock \emph{International Journal of Behavioral Medicine}, 2020.

\bibitem[Stach et~al.(2020)Stach, Kraft, Probst, Messner, Terhorst, Baumeister,
  Schickler, Reichert, Sander, and Pryss]{stach2020}
Michael Stach, Robin Kraft, Thomas Probst, Eva-Maria Messner, Yannik Terhorst,
  Harald Baumeister, Marc Schickler, Manfred Reichert, Lasse~Bosse Sander, and
  R{\"u}diger Pryss.
\newblock {Mobile Health App Database - A Repository for Quality Ratings of
  mHealth Apps}.
\newblock In \emph{{2020 IEEE 33rd International Symposium on Computer-Based
  Medical Systems (CBMS)}}. IEEE, 2020.

\bibitem[Terhorst et~al.(2018)Terhorst, Rathner, Baumeister, and
  Sander]{terhorst2018}
Yannik Terhorst, Eva-Maria Rathner, Harald Baumeister, and Lasse~Bosse Sander.
\newblock "hilfe aus dem app-store": Eine systematische Übersichtsarbeit
  undevaluation von apps zur anwendung bei depressionen.
\newblock \emph{Verhaltenstherapie}, 2018.

\bibitem[Terhorst et~al.(2021)Terhorst, Messner, Schultchen, Paganini,
  Portenhauser, Eder, Bauer, Papenhoff, Baumeister, and Sander]{terhorst2021}
Yannik Terhorst, Eva-Maria Messner, Dana Schultchen, Sarah Paganini, Alexandra
  Portenhauser, Anna-Sophia Eder, Melanie Bauer, Mike Papenhoff, Harald
  Baumeister, and Lasse~Bosse Sander.
\newblock Systematic evaluation of content and quality of english and german
  pain apps in european app stores.
\newblock \emph{Internet Interventions}, 24:\penalty0 100376, 2021.

\bibitem[AppRadar(2021)]{AppRadar}
AppRadar.
\newblock \url{https://appradar.com/de}, 2021.
\newblock Accessed: 2021-04-01.

\bibitem[Cor(2021{\natexlab{a}})]{CoronaC}
{Corona Check}, 2021{\natexlab{a}}.
\newblock URL \url{https://www.coronacheck.science/en/}.

\bibitem[Cor(2021{\natexlab{b}})]{CoronaH}
{Corona Health}, 2021{\natexlab{b}}.
\newblock URL \url{https://www.corona-health.net/en/}.

\bibitem[Kraft et~al.(2020)Kraft, Schlee, Stach, Reichert, Langguth,
  Baumeister, Probst, Hannemann, and Pryss]{kraft2020combining}
Robin Kraft, Winfried Schlee, Michael Stach, Manfred Reichert, Berthold
  Langguth, Harald Baumeister, Thomas Probst, Ronny Hannemann, and R{\"u}diger
  Pryss.
\newblock Combining mobile crowdsensing and ecological momentary assessments in
  the healthcare domain.
\newblock \emph{Frontiers in neuroscience}, 14:\penalty0 164, 2020.

\bibitem[Pryss(2019)]{pryss2019mobile}
R{\"u}diger Pryss.
\newblock Mobile crowdsensing in healthcare scenarios: taxonomy, conceptual
  pillars, smart mobile crowdsensing services.
\newblock In \emph{Digital Phenotyping and Mobile Sensing}, pages 221--234.
  Springer, 2019.

\bibitem[169(2016)]{CouncilDirective}
OJ~L 169.
\newblock {12.7. 1993. Council Directive 93/42/EEC of 14 June 1993 concerning
  medical devices}.
\newblock
  \url{https://eur-lex.europa.eu/legal-content/EN/TXT/?uri=CELEX:31993L0042},
  2016.
\newblock Accessed: 2021-03-26.

\bibitem[117(2018)]{Regulation}
OJ~L 117.
\newblock {Regulation (EU) 2017/745 of the European Parliament and of the
  Council of 5 {April} 2017 on medical devices, amending Directive 2001/83/EC,
  Regulation (EC) No 178/2002 and Regulation (EC) No 1223/2009 and repealing
  Council Directives 90/385/EEC and 93/42/EEC. 2017 May 05.}
\newblock
  \url{https://eur-lex.europa.eu/legal-content/EN/TXT/?uri=CELEX\%3A32017R0745},
  2018.
\newblock Accessed: 2021-03-26.

\bibitem[Keutzer and Simonsson(2020)]{MDRKeutzer}
Lina Keutzer and Ulrika~SH Simonsson.
\newblock {Medical Device Apps: An Introduction to Regulatory Affairs for
  Developers}.
\newblock \emph{JMIR Mhealth Uhealth}, 8\penalty0 (6):\penalty0 e17567, Jun
  2020.
\newblock \doi{10.2196/17567}.

\bibitem[Lang(2017)]{MDD}
Michael Lang.
\newblock {Heart Rate Monitoring Apps: Information for Engineers and
  Researchers About the New European Medical Devices Regulation 2017/745}.
\newblock \emph{JMIR Biomed Eng}, 2\penalty0 (1), Aug 2017.
\newblock \doi{10.2196/biomedeng.8179}.

\bibitem[Trektere et~al.(2016)Trektere, McCaffery, Lepmets, and
  Barry]{MDevSpiceTrektere}
Kitija Trektere, Fergal McCaffery, Marion Lepmets, and Grainne Barry.
\newblock {Tailoring MDevSPICE® for Mobile Medical Apps}.
\newblock In \emph{Proceedings of the International Conference on Software and
  Systems Process}. Association for Computing Machinery, 2016.
\newblock \doi{10.1145/2904354.2904361}.

\bibitem[Trektere et~al.(2017)Trektere, Regan, Caffery, Flood, Lepmets, and
  Barry]{TracebilityTrektere}
Kitija Trektere, Gilbert Regan, Fergal~Mc Caffery, Derek Flood, Marion Lepmets,
  and Grainne Barry.
\newblock Mobile medical app development with a focus on traceability.
\newblock \emph{Journal of Software: Evolution and Process}, 29\penalty0
  (11):\penalty0 e1861, 2017.
\newblock \doi{https://doi.org/10.1002/smr.1861}.

\bibitem[IEC(2015)]{IEC62304}
{IEC62304:2006/AMD1:2015 Amendment 1 - Medical device software - Software life
  cycle processes}.
\newblock \url{https://webstore.iec.ch/publication/22790}, 2015.
\newblock Accessed: 2021-03-30.

\bibitem[GAM(2008)]{GAMP5}
Gamp 5 guide.
\newblock \url{https://www.ispe.org/publications/guidance-documents/gamp-5},
  2008.
\newblock Accessed: 2021-03-30.

\bibitem[Health et~al.(2011)Health, Food, Drug~Administration,
  Radiological~Health, and Research]{FDA}
U.S. Department~Of Health, Human~Services Food, Center for~Devices
  Drug~Administration, Center for Biologics~Evaluation Radiological~Health, and
  Research.
\newblock {General Principles of Software Validation; Final Guidance for
  Industry and FDA Staff}.
\newblock
  \url{https://www.fda.gov/regulatory-information/search-fda-guidance-documents/general-principles-software-validation},
  2011.
\newblock Accessed: 2021-03-30.

\bibitem[Convention(2018)]{PIC}
Pharmaceutical~Inspection Convention.
\newblock Guide to good manufacturing practice for medicinal products annexes.
\newblock \url{https://picscheme.org/docview/2470}, 2018.
\newblock Accessed: 2021-03-30.

\bibitem[IEC(2016)]{IEC82304}
{IEC 82304-1:2016 Health software — Part 1: General requirements for product
  safety}.
\newblock \url{https://www.iso.org/standard/59543.html}, 2016.
\newblock Accessed: 2021-03-30.

\bibitem[Agency(2015)]{ICH}
European~Medicines Agency.
\newblock {ICH guideline Q9 on quality risk management}.
\newblock
  \url{https://www.ema.europa.eu/en/documents/scientific-guideline/international-conference-harmonisation-technical-requirements-registration-pharmaceuticals-human-use_en-3.pdf},
  2015.
\newblock Accessed: 2021-03-30.

\bibitem[Vogel et~al.(2021 (in print))Vogel, Pryss, Schobel, Schlee, and
  Beierle]{Vogel2021MobileSoft}
Carsten Vogel, R{\"u}diger Pryss, Johannes Schobel, Winfried Schlee, and Felix
  Beierle.
\newblock {Developing Apps for Researching the COVID-19 Pandemic with the
  TrackYourHealth Platform}.
\newblock In \emph{2021 {{IEEE}}/{{ACM}} 8th {{International Conference}} on
  {{Mobile Software Engineering}} and {{Systems}} ({{MOBILESoft}})}. IEEE, 2021
  (in print).

\end{thebibliography}

\end{document}